\documentclass[preprint,showpacs,aps]{revtex4}

\usepackage{graphicx}
\usepackage{dcolumn}
\usepackage{bm}

\begin{document}

\begin{titlepage}

\title{Strain Control of Magnetism in Transition-Metal-Atom Decorated Graphene}

\author{Bing Huang$^{1}$\footnote{E-mail: Bing.Huang@nrel.gov}, Jaejun
  Yu$^{1,2}$\footnote{E-mail: jyu@snu.ac.kr}, and Su-Huai Wei$^{1}$\footnote{E-mail: swei@nrel.gov}}

\affiliation{$^1$National Renewable Energy Laboratory, 1617 Cole Boulevard, Golden, CO 80401, USA}
\affiliation{$^2$Center for Strongly Correlated Materials Research, Department of Physics and Astronomy,
Seoul National University, Seoul 151-747, Korea}

\date{\today}

\begin{abstract}
  We report a strain-controlled tuning of magnetism in transition-metal-atom-decorated graphene.
  Our first-principles calculations demonstrate that strain can lead to a sudden change
  in the magnetic configuration of a transition metal (TM) adatom and the local
  atomic structure in the surrounding graphene layer, which have a dramatic effect on
  the effective exchange coupling between neighboring TM atoms. A
  strong spin-dependent hybridization between TM $d$ and graphene $\pi$
  orbital states, derived from the orbital selection rule of the local lattice
  symmetry, is responsible for the determination of the local electronic
  and magnetic structure. Our results indicate that the strain can be an effective way
  to control the magnetism of atomic-scale
  nanostructures, where the reliable control of their magnetic states is a key step for the
  future spintronic applications.
\end{abstract}


\maketitle

\vspace{2mm}

\end{titlepage}

A reliable control of magnetic states is central to the use of
magnetic nanostructures in future spintronics and quantum
information devices\cite{S. A. Wolf-2001}. Although it is shown that
magnetism could be generally modulated by external magnetic \cite{O.
Kahn-1998, S. A. Wolf-2001, C. F. Hirjibehedin-2006} or electric
\cite{M. Weisheit-2007} fields, it is desirable to find an alternative
scheme to control the magnetism for various spintronic
applications. Recently, graphene has gathered tremendous attention
due to its unique electronic and mechanical properties for nanoscale
electronics\cite{A. K. Geim-2009}. As a candidate material for
spintronic devices, transition-metal-atom-decorated graphene
(denoted as TM-graphene hereafter) has been studied extensively in
theory\cite{B. Uchoa-2008, A. V. Krasheninnikov-2009, B.
Uchoa-2011} and experiment\cite{K. Pi-2009, V. W. Brar-2010,
Cretu-2010}, manifesting some remarkable electronic and magnetic
behaviors. Thus, developing a novel method to tune the magnetism of
TM-graphene system is quite urgent for future spintronics applications.

Graphene is the thinnest material ever synthesized\cite{A. K.
Geim-2009}. While it is one of the strongest materials ever measured
experimentally\cite{C. Lee-2008, K. S. Kim-2009, Levy-2010, W.
Bao-2009}, it has been shown that it can sustain elastic
deformations as large as 25\%. In this Letter, we predict that the
strain is an effective way to control the magnetic properties of
TM-graphene systems. As for the graphene layers, we considered a
pristine graphene (PG) layer as well as graphene layers with defects
of a single vacancy (SV) or a double vacancy (DV). It has been shown
that vacancies in graphene can be created by electron
irradiation\cite{A. Hashimoto-2004, A. V. Krasheninnikov-2007} or
ion bombardment\cite{M. M. Ugeda-2010}. Our results for the
Mn-atom-decorated graphene demonstrate that the strain can induce a
sudden change of the local atomic structure of graphene around a Mn
atom and the spin state of Mn atom, which can lead to a striking
change of the effective exchange coupling between neighboring Mn
magnetic moments. A strong spin-dependent hybridization between Mn
$d$ and graphene $\pi$ orbital or dangling bond states is
responsible for the determination of the local electronic and
magnetic structure. We also showed that the strain-controlled
magnetism in Mn-graphene is a general phenomenon happening in
different TM-graphene systems.

All the density-functional-theory (DFT) calculations were performed
using VASP code\cite{VASP}. We used the projector augmented wave
(PAW) potentials and the generalized gradient approximation with the
Perdew-Burk-Ernzerhof (PBE) functional to describe the core
electrons and the exchange-correlation energy, respectively, which
were proved to describe well the TM-graphene systems\cite{A. V.
Krasheninnikov-2009, K. T. Chan-2008, Sevincli-2008}. A 7$\times$7
graphene supercell (98 C atoms) was used in our calculations,
approaching the single defect limit ($~$ 1\% defect concentration).
A 12$\times$12 supercell was used to calculate the interaction of TM
atoms on graphene. A $\Gamma$-centered 6$\times$6$\times$1
\textbf{k}-point sampling was used for Brillouin-zone integration.
Moreover GGA+$U$ (with the on-site Hubbard $U$ correction)
calculations with $U$ = 5 eV were performed to confirm the GGA
results of the Mn-graphene system. We verified that both GGA and
GGA+$U$ results are consistent with each other.

\begin{figure}[tbp]
\includegraphics[width=12.0cm]{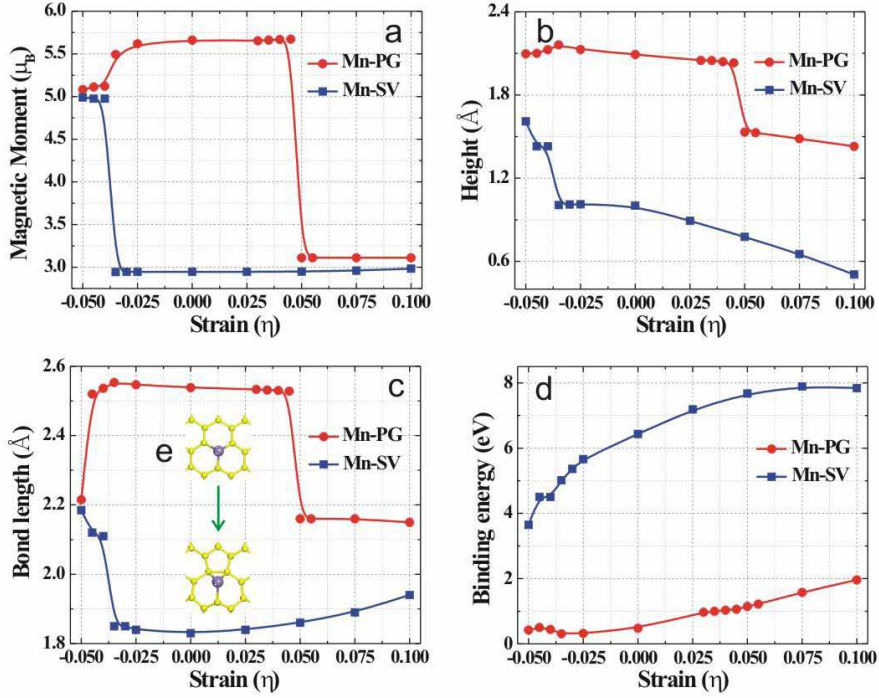}
\caption{(a) Magnetic Moments of Mn atoms adsorption on prefect
graphene (PG) and defected graphene with a single-vacancy (SV) as a
function of strain. (b) The heights between Mn atoms and the nearest
C atoms as a function of strain for Mn-PG and Mn-SV systems. (c) The
average bond lengths between Mn atoms and the nearest C atoms as a
function of strain for Mn-PG and Mn-SV systems. (d) The binding
energies of Mn atoms on graphene as a function of strain for Mn-PG
and Mn-SV systems. (e) The local structure transition for Mn-SV
system from $\eta$ = 0 (up structure) to $\eta$ = -0.04 (down
structure).}
\end{figure}

We take the case of Mn atom adsorption on graphene as a primary example to
demonstrate the strain effect on the TM-graphene system. The
calculated strain-dependent magnetic moments are shown in Fig.~1a. A biaxial
strain $\eta$ is defined as $\eta=\Delta a/a$,
where $a$ is the lattice constant of free standing graphene. A
positive (negative) $\eta$ means the TM-graphene system is under
tension (compression). For PG without vacancies, Mn atom prefers to
stay at the hollow site (above the carbon hexagon center) of PG and
the total magnetic moment of Mn atom is 5.7 $\mu_B$, indicating that
Mn atom is in a high-spin state. The height and average bond length
between Mn atom and its nearest C atoms is 2.09 {\AA} and 2.54 {\AA},
respectively, as shown in Fig. 1b and 1c. The binding energy of Mn on
PG is about 0.45 eV, as shown in Fig. 1d. When a biaxial
tensile strain $\eta$ is applied for Mn-PG, interestingly,
the magnetic ground state changes from high-spin to low-spin
($\sim$ 3.1 $\mu_B$) abruptly at the critical strain of $\eta$ = 0.05, as
shown in Fig. 1a. Accompanying the magnetic transition, the height and
average bond length between Mn atom and its nearest C atoms decrease
sharply to 1.50 and 2.16 \AA, respectively, as shown in Fig. 1b and 1c.
The binding energy of Mn atom of Mn-PG increases to 1.15 eV when
$\eta$ = 0.05. In contrast to the sharp transitions of spin-states
and local adsorption structures, the binding energy of Mn atom on PG
increases continuously as a function of $\eta$, as shown in Fig.
1d.

TM atoms on PG have low migration barriers so that it may be mobile at
room temperature \cite{A. V. Krasheninnikov-2009, Cretu-2010, K. T.
  Chan-2008, Sevincli-2008}. However, our climbing-image
nudged-elastic-band calculations\cite{NEB} show that the migration
barrier for the Mn atom on PG increases from 0.36 eV to 0.89 eV
under a tension of $\eta=0.10$, giving rise to a decrease of more
than 8 orders of magnitude in the diffusion coefficient at room
temperature, which implies that the migration can be suppressed
significantly. It also indicates that an adequate variation of
strain can be used to control the patterning of TM atoms on
graphene. Furthermore, it is known that the mobility of the TM atom
becomes markedly reduced in the presence of vacancies. For instance,
the migration barriers for TM atoms on vacancies are larger than 3
eV\cite{A. V. Krasheninnikov-2009}.

A similar magnetic transition between high-spin and low-spin states is
also found for Mn atoms adsorbed near the SV sites in graphene  (Mn-SV).
Different from the Mn-PG case,
the unstrained Mn-SV is in a low-spin state with the magnetic
moment of $\sim$ 3 $\mu_B$, as shown in Fig. 1a. The local
adsorption structures of the Mn atoms are shown in Fig. 1e.
The binding energy between Mn atom and
vacancy is 6.4 eV, which is much larger compared to the Mn-PG case,
indicating that the Mn-SV system is quite stable. This is in agreement
with previous results\cite{A. V. Krasheninnikov-2009}.
Interestingly, a compression of $\eta = -0.04$ can convert the magnetic
ground state from low-spin ($\sim$ 3 $\mu_B$) to high-spin
($\sim$ 5 $\mu_B$) for Mn-SV. Under the compression the
graphene layer produces a spontaneous rippling with the formation of a C-C
dimer bond near SV, as shown
in Fig.~1e. As results, the height and the average bond length between
Mn and its nearest C atoms increase significantly under
compression, which weakens the binding between Mn atom and
vacancy. In general, we find that tensile (compressive) strain can
enhance (reduce) the binding strength between TM atom and graphene
(Fig. 1d), which is consistent with an experimental
observation\cite{Cretu-2010}.

\begin{figure*}[tbp]
\includegraphics[width=14.0cm]{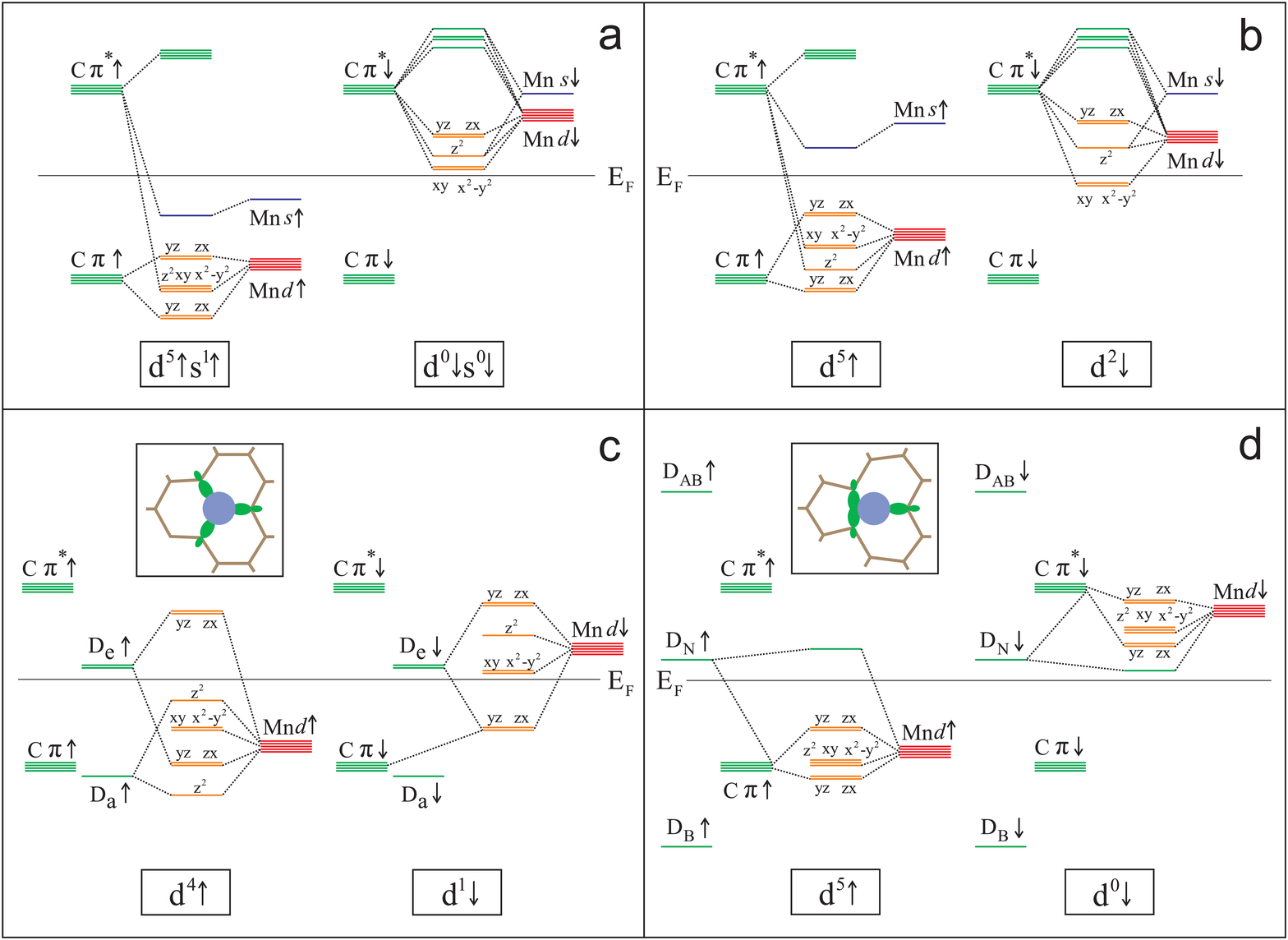}
\caption{Schematic drawings of the energy diagrams of Mn atom and graphene
  electronic states near the Fermi level ($E_{F}$) for (a) Mn atom on PG
  under $\eta$ = 0, (b) Mn atom on PG under $\eta$ = 0.05, (c) Mn atom on
  graphene with SV under $\eta$ = 0, and (d) Mn atom on graphene with SV
  under $\eta$ = -0.04. The local electronic structure of graphene is
  represented by the occupied $\pi$ and unoccupied $\pi^{\star}$ levels
  arising from C $p_{\pi}$ bands, which touch at the K point in the orignal Brillouin
  zone of graphene.  In (c), three dangling bond orbitals of C atoms near
  SV form one bonding state, labelled as $D_a$, and two degenerate
  non-bonding states, labelled as $D_{e}$. While the
  bonding $D_{a}$ level is close to $\pi$, the non-bonding $D_{e}$
  level is located just above $E_{F}$. In (d), the distortion and formation
  of C-C pair gives rise to the bonding and anti-bonding states $D_{B}$
  and $D_{AB}$, which stay far away from the Fermi level, and one non-bonding
  dangling bond state $D_{N}$ just above $E_{F}$.}
\end{figure*}

It is interesting to observe that the electronic and magnetic
structures of the Mn atom depend strongly on the strain of graphene
layer, thereby leading to the strain-induced sudden change of the magnetic
ground state. The strain affects the local geometry of carbon atoms
around the Mn atom in two ways. It changes the height of Mn atom
above the graphene layer and generates a ripple structure under
the compressive strain. The change of the local environment, which
determines the distance and symmetry of C $p$ orbitals with respect to
that of Mn $d$ orbitals, can lead to a dramatic change in the magnetic
$d$ orbitals due to the strong spin-dependent
hybridization between Mn $d$ and C $\pi$ orbital states.

The isolated Mn atom
has $s^{1\uparrow} d^{5\uparrow} s^{1\downarrow} d^{0\downarrow}$ configuration with
a spin exchange splitting around 4 eV. Since the Mn atom
sits over the hollow site, the energy levels of 3$d$ orbitals are
broadened by the ligand field from the hybridization between Mn 3$d$ and C
$p_{\pi}$ orbitals of the hexagon carbon atoms at the hollow site. Moreover,
because the Mn spin-down energy level is above the graphene Fermi energy, one
electron is transferred from Mn to the graphene layer. This is consistent
with the observation that the magnetic configuration of the Mn
atom in the Mn-PG case can be ascribed to $d^{5\uparrow}
s^{1\uparrow} d^{0\downarrow} s^{0\downarrow}$ with the calculated magnetic moment of
5.7 $\mu_{\mathrm{B}}$ shown in Fig.~1a.

In the Mn-PG system, the local geometry of the Mn adatom above the
center of the carbon hexagon has the $C_{6v}$ symmetry. While Mn
$d_{yz,zx}$ and $d_{xy,x^{2}-y^{2}}$ orbitals belong to the $E_{1}$
and $E_{2}$ representations of $C_{6v}$, respectively, the $p_{\pi}$
electrons near the Fermi level ($E_{F}$) can be also classified to
$A_{1}$, $E_{1}$ (corresponding to the $\pi$ state) and $B_{2}$,
$E_{2}$ ($\pi^{\star}$ state). From the orbital selection rule, for
example, the C $\pi^{\star}$ state couples most strongly with Mn
$d_{xy,x^{2}-y^{2}}$ orbitals because both of them belong to the
same symmetry representation. Due to the exchange splitting of Mn
$d$ states, however, the spin-up Mn $d^{\uparrow}$ levels are close
to the occupied C $\pi$ states and the spin-down Mn $d^{\downarrow}$
levels are located near the unoccupied C $\pi^{\star}$ states.
Consequently, the proximity between Mn $d^{\downarrow}$ and C
$\pi^{\star}$ states makes the
$d_{xy,x^{2}-y^{2}}^{\downarrow}$-$\pi^{\star}$ hybridization much
stronger so that the unoccupied $d_{xy,x^{2}-y^{2}}^{\downarrow}$
state is pushed down close to $E_{F}$, as illustrated in Fig.~2a.

When the graphene is under a tensile strain, the height of Mn atom lowers
and the bond lengths between Mn and C atoms decreases (Figs. 1b and
1c). At $\eta \sim 0.05$ the change becomes abrupt. This is because the reduced bond
length with the lower height of Mn increases the hybridization
strength between Mn $d_{xy,x^2-y^2,z^2}$ and C $\pi^{\star}$ states so
that the $d_{xy,x^2-y^2}$ spin-down state becomes
occupied. Since the Mn $d_{xy,x^2-y^2}$ can hold two electrons,
consequently, the occupation of the spin-down Mn $d_{xy,x^2-y^2}$ orbital reduces the
total magnetic moment and the exchange splitting as well. Therefore the
final magnetic configuration of Mn-PG for $\eta \geq 0.05$ becomes
$d^{5\uparrow} d^{2\downarrow}$ with the magnetic moment of 3
$\mu_{\mathrm{B}}$.

In the case of Mn-SV, one has to consider the role of C dangling bond
states arising from the carbon vacancy as illustrated in
Fig.~2c and 2d. At $\eta=0.0$, the
new structure can form without a significant lattice distortion. When a carbon vacancy is present
within the graphene layer, the Mn atom is much closer to the graphene
plane and the C-Mn bond length is much shorter. In this case, the main
hybridization is between the Mn $d$ state and the C dangling bond states
from the neighboring C atoms. Similar to the case of Mn atom on PG,
the coupling between the dangling bond state $D_{a}$ and $D_{e}$ and the
Mn $d$ states depends on the symmetries of the states.
For example, the largest coupling occurs between Mn
$d_{yz,zx}$ and C $D_{e}$ states, whereas Mn $d_{z^{2}}$
couples mostly to $D_{a}$. Here we have a
similar spin-dependent hybridization effect due to the proximity of Mn
$d^{\uparrow}$ to $D_{a}$ and Mn $d^{\downarrow}$ to $D_{e}$ levels,
respectively. Due to the strong $d_{yz,zx}$-$D_{e}$ coupling, the
bonding-antibonding separation of the $d_{yz,zx}$-$D_{e}$ hybrid
states becomes much stronger than in the Mn-PG case, so that the
antibonding Mn $d_{yz,zx}^\uparrow$-derived state becomes unoccupied,
whereas the bonding Mn $d_{yz,zx}^\downarrow$-derived state is
occupied as shown in Fig.~2c. Although the occupied bonding states are doubly degenerate,
its covalent character reduces the effective spin moment
close to 1 $\mu_{\mathrm{B}}$. Thus, as illustrated in Fig. 2c, the
magnetic configuration for the case of Mn-SV without strain becomes
$d^{4\uparrow}$ and $d^{1\downarrow}$ (the Mn s orbital are in higher
energy due to the charge transfer form Mn to graphene), which corresponds
to the calculated magnetic moment of 3 $\mu_{\mathrm{B}}$ shown in
Fig. 1a.

To achieve the high-spin state in Mn-SV, as we have learned from the
Mn-PG case, we should increase the Mn-C distance to reduce the
hybridization between the Mn d and C $p_{\pi}$ state. Ironically,
this should be done by compressive strain. Indeed, we find that for
$\eta \leq -0.04$, the hight of Mn increases drastically, so is the
Mn-C bond length as shown in Fig. 1c. Moreover, the graphene layer
around the SV region becomes strongly rippled and two of the three
carbon atoms around the vacancy site form a pair (Fig.~1e and 2d).
The C dangling bond states now becomes $D_{B}$, $D_{N}$, and
$D_{AB}$ (Fig. 2d). Because of the elongated structure, the
hybridization between the $D_{N}$ and Mn $d$ orbitals becomes much
reduced in the rippled structure so that the Mn atom maintains its
$d^{5\uparrow} d^{0\downarrow}$ configuration with the minimal
$d$-level broadening for both spin-up and spin-down channels.
Therefore, the spin transition of Mn-SV system indicates that the
rippled graphene structures (induced by compression) play an
important role in the spin transition from low-spin to high-spin
state. It is quite encouraging to see that the rippling of graphene
layers already can be controlled by substrates or thermal expansion
in experiments\cite{A. K. Geim-2009, W. Bao-2009}.

\begin{table}
\caption{\label{tab:table2} The energy differences between
ferromagnetic (FM) state and antiferromagnetic (AFM) state
($E_{\mathrm{AFM}}-E_{\mathrm{FM}}$) for Mn atoms on pristine
graphene as a functional of strain $\eta$. The Mn-Mn distance is fixed at $\sim$ 9 {\AA}. The corresponding Curie temperature $T_{c}$ are
also shown.}
\begin{ruledtabular}
\begin{tabular}{cccccc}
$\eta$ & 0.000 & 0.025 & 0.050 & 0.075 & 0.100
\\[2pt]
\hline
\ $E_{\mathrm{AFM}}-E_{\mathrm{FM}}$ (meV) & -25.1  & -9.3 & 29.4 & 23.5 & 17.6  \\[2pt]
\ $T_{c}$ (K) & 129  & 48 & 151 & 121 & 90 \\[2pt]
\end{tabular}
\end{ruledtabular}
\end{table}

The interaction of localized Mn moments is evaluated by calculating
the energy difference ($E_{\mathrm{AFM}}-E_{\mathrm{FM}}$) between
ferromagnetic (FM) state and antiferromagnetic (AFM) state. Taking
Mn-PG system as an example, Table I shows the calculated
$E_{\mathrm{AFM}}-E_{\mathrm{FM}}$ per Mn-Mn pair as a functional of
$\eta$. The Mn-Mn distance is set to $\sim$ 9 \AA, which is a
reasonable doping concentration in practice. The calculated
$E_{\mathrm{AFM}}-E_{\mathrm{FM}}$ can be related to the Curie
temperature $T_{c}$ through an effective Heisenberg model with an
interaction $J = (E_{\mathrm{AFM}}-E_{\mathrm{FM}})/2$.  The
positive energies in Table I indicate FM spin alignment at low
temperature while negative values are AFM ones. Without strain, Mn
atoms on graphene prefer the AFM coupling in agreement with previous
results\cite{Sevincli-2008}. This is consistent with the fact that
the spin-up $d$ orbital is fully occupied and spin-down $d$ orbitals
are fully empty. The spin hopping is allowed in the AFM arrangement
but not allowed in the FM configuration, resulting in a lower energy
of AFM state\cite{Dalpian-2006}. Interestingly, strain dramatically
changes the spin coupling between Mn atoms on graphene when the Mn
spin-down $d$ orbitals become partially occupied and spin hopping is
allowed for FM arrangement. For $\eta$ $\geq$ 0.05, the magnetic
ordering converts from AFM to FM, consistent with the expectation of
the band coupling model description of the magnetic
interactions\cite{Dalpian-2006}. This result strongly demonstrates
that the spin exchange strength and magnetic ordering of a
TM-graphene system could be switched by simply using a strain. The
Curie temperature, $T_{c}$, can be roughly estimated by using the
formula $T_c = 0.44JS(S + 1)/k_B$ which correspondings to a quantum
Monte-Carlo (QMC) study of a quasi two-dimensional Heisenberg
ferromagnet on a square lattice\cite{C. Yasuda-2005}. The results
are shown in Table I. For example, a value of $T_c = 151$ K is
obtained when $\eta$ = 0.05.

\begin{figure*}[tbp]
\includegraphics[width=16.0cm]{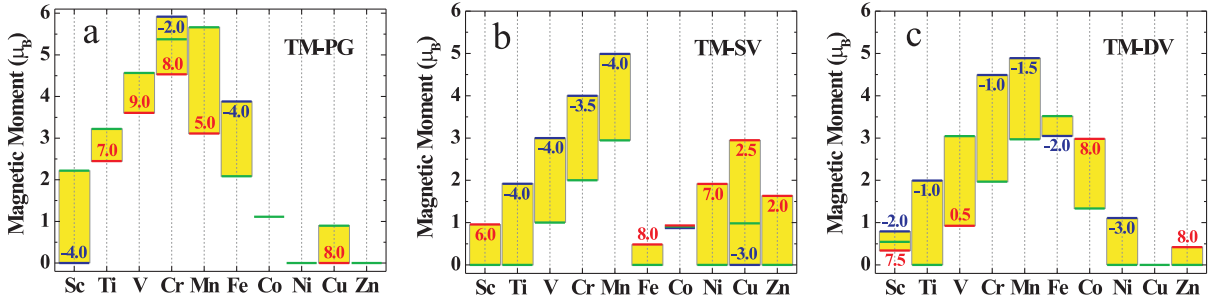}
\caption{The range of magnetic moments of TM atoms adsorption on
graphene (a) without vacancy, (b) with SV, (c) with DV under -0.05
$\leq$ $\eta$ $\leq$ 0.10. The intrinsic magnetic moments of
TM-graphene without strain are shown as green line, while the
maximum and minimum magnetic moments of TM-graphene caused by
compressive (or tensible) strain (in unit of \%) are shown by blue
(or red) line.}
\end{figure*}

Not only for the Mn atom on graphene, the strain control of the
magnetism generally exists for almost all the TM atoms from Sc to Zn
on graphene. The range of magnetic moments of TM-graphene systems
under realistic strain -0.05 $\leq$ $\eta$ $\leq$ 0.10 are shown in
Fig.~3. The transition of magnetic states for TM-PG systems under
strain could be either continuous (e.g., V, Ti, and Cr) or abrupt
(e.g., Sc, Mn, Fe, and Cu), as shown in Fig.~3a, which is mainly due
to different strain dependent ligand splitting and hybridization.
Interestingly, strain can also control the adsorption sites of TM
atom, \emph{e.g.}, the adsorption sites of Cu (Sc and Fe) on
graphene converts from C-C bridge (hollow) site to hollow (C-C
bridge) site at $\eta$ $\geq$ 0.08 ($\eta$ $\leq$ -0.04).

The spin transition is even more remarkable in the TM-SV cases, as
shown in Fig.~3b. For example, a compressive strain of -0.04, -0.04,
and -0.035 could transfer Ti-, V- and Cr-SVs from low-spin to
high-spin states, respectively, similar to Mn-SV. A tensile strain
of 0.025, 0.06, and 0.07 could transfer Cu-, Sc-, and Ni-SV from
low-spin to high-spin states, respectively. It should be noticed
that at least 66\% (50\%) of total magnetic moment of Cu-SV (Ni-SV)
system origins from the nearest C atoms around vacancy. We have also
studied the cases of TM-DV, since the stability of TM atoms adsorbed
on DV is comparable to that of SV\cite{A. V. Krasheninnikov-2009}.
The spin transition in TM-DV system is similar to that of TM-SV but
with a smaller critical $\eta$ for spin transition, as shown in
Fig.~3c. Especially, a compressive strain of -0.01, -0.01, and -0.015
could convert the spin-states of Ti-, Cr-, and Mn-DV from low-spin
to high-spin, respectively. Surprisingly, a very small tension of
0.005 could transfer V-DV from high-spin to low-spin state.

In conclusion, a concept of controlling the magnetism of TM-graphene
by simply using a strain is reported. Our calculations demonstrate
that the strain-induced change of the local atomic structure of
graphene around a TM atom makes a dramatic effect on both the spin
state of the TM atom and the exchange coupling between neighboring
TM magnetic moments. A strong spin-dependent hybridization between
TM $d$ and graphene $p$ orbital states, derived from the orbital
selection of the local lattice symmetry, is responsible for the
determination of the local electronic and magnetic structure. Our
results indicate that the strain is available to control the
magnetism of nanoscale materials for spintronics.

The work at NREL was supported by the U.S. Department of Energy
under Contract No. DE-AC36-08GO28308. J.Y. acknowledges the support
by the National Research Foundation of Korea through the ARP (No.
R17-2008-033-01000-0).

\newpage

\end{document}